\documentclass[conference]{IEEEtran}
\ifCLASSINFOpdf
 \usepackage[pdftex]{graphicx}
\else
\fi
\usepackage[lined,algoruled,noline]{algorithm2e}
\usepackage{amsmath}
\usepackage{times}
\usepackage{pifont}
\usepackage{txfonts}

\usepackage{tikz}
\usepackage{textcomp}
\usepackage{hyperref}
\usepackage{lipsum}

\newcommand\copyrighttext{%
  \footnotesize \textcopyright 2015 IEEE. Personal use of this material is permitted. Permission from IEEE must be obtained for all other uses, in any current or future media, including reprinting/republishing this material for advertising or promotional purposes, creating new collective works, for resale or redistribution to servers or lists, or reuse of any copyrighted component of this work in other works.
  DOI: 10.1109/ICARA.2015.7081159}
\newcommand\copyrightnotice{%
\begin{tikzpicture}[remember picture,overlay]
\node[anchor=south,yshift=10pt] at (current page.south) {\fbox{\parbox{\dimexpr\textwidth-\fboxsep-\fboxrule\relax}{\copyrighttext}}};
\end{tikzpicture}%
}

\makeatletter
\newcommand{\removelatexerror}{\let\@latex@error\@gobble}
\makeatother

\begin{document}
%\title{Validation of parameter inference and uncertainty reporting from sensor data}
\title{Automated weighing by sequential inference in dynamic environments}

\author{\IEEEauthorblockN{A. D. Martin and T. C. A. Molteno}
\IEEEauthorblockA{
Department of Physics\\ 
University of Otago\\ 
Dunedin, 9016, New Zealand\\
Email: amartin@elec.ac.nz
}
}


%Telephone: (800) 555--1212, Fax: (888) 555--1212}


\maketitle

\copyrightnotice

\begin{abstract}
%\boldmath
We demonstrate sequential mass inference of a suspended bag of milk powder from simulated measurements of the vertical force component at the pivot while the bag is being filled. We compare the predictions of various sequential inference methods both with and without a physics model to capture the system dynamics. We find that non-augmented and augmented-state unscented Kalman filters (UKFs) in conjunction with a physics model of a pendulum of varying mass and length provide rapid and accurate predictions of the milk powder mass as a function of time. The UKFs outperform the other method tested - a particle filter. Moreover, inference methods which incorporate a physics model outperform equivalent algorithms which do not.
\end{abstract}


% \fi
\IEEEpeerreviewmaketitle

\section{Introduction}\label{Sec:Intro}
Sequential inference has been used for measurement and control in many contexts including vehicle navigation \cite{Julier:2004}, target tracking \cite{Costa:1994} and chemical process plant control \cite{Prasad:2000}. Sequential inference algorithms take a time-series of noisy measurements of a system, and produce increasingly accurate estimates of the system parameters or state variables in the form of a posterior distribution. We adapt sequential inference algorithms for use in automated weighing systems, where the system under consideration exhibits dynamics according to physical laws. Such methods provide dynamically updated estimates of mass, along with estimates of its uncertainty, which will be useful for the control of automated weighing systems. 

We use the example of a milk powder bagging system, where the bag is suspended from a point at its top from which it may swing under the influence of gravity. The bag is gradually filled with milk powder while measurements of the vertical force component are made at discrete times. We test sequential inference algorithms by simulating such measurements, taking into account both process and measurement noise, and run the algorithms on the simulated data-set. We analyse the predictions for accuracy, precision and speed of inference, and draw conclusions about the most suitable algorithm. We stress the benefits of considering the underlying physics when designing sensors and control systems.

\section{Methods}\label{Sec:Methods}
\subsection{Physics model}\label{Sec:Physics}
We model the bag-filling system as a pendulum with mass increasing at rate $\dot{m}$ (as the bag fills), and with effective length $l = L-x_{\mbox{\scriptsize com}}$, where $x_{\mbox{\scriptsize com}} = m/(2\rho A)$,  $\rho$ is the density of the milk powder and $A$ is the cross-sectional area of the bag. The rate $\dot{m}$ is assumed to vary randomly (see below). Figure \ref{Fig:schematic} (a) shows a schematic of the system.
The equations of motion for the pendulum are simply:
\begin{align}
\dot{\theta} = \omega,\label{Eq:theta}\\ 
\dot{\omega} = -g\sin\theta/l,\label{Eq:omega}
\end{align} where $\theta$ is the angle of the pendulum, and the effective pendulum length $l$ is not constant but decreases as the milk powder's centre of mass rises.
\begin{figure}
%\centering
\includegraphics[width=\columnwidth]{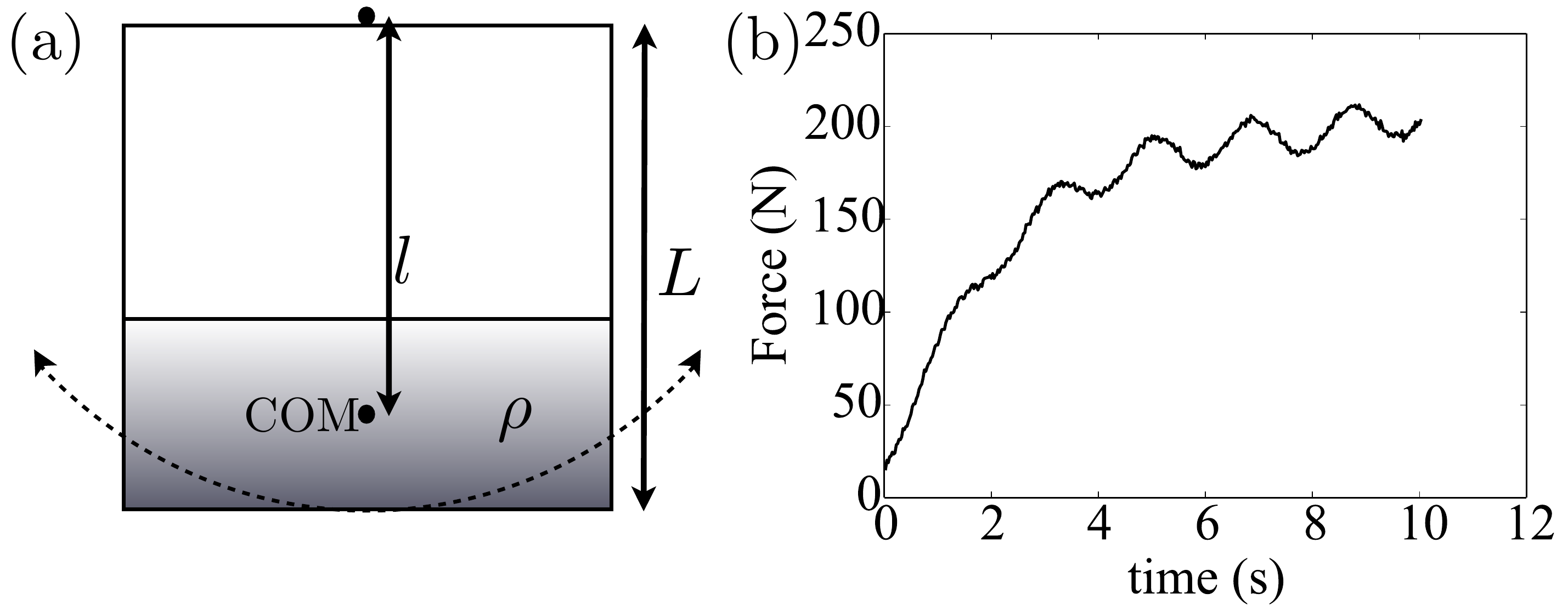}
\caption{(a) Schematic of the system. The box is a model of the bag of length $L$, which pivots at the indicated point at its top. The cross-sectional area is $A$ (not indicated on the figure). The milk powder is represented by the shaded region, and has density $\rho$, and its centre of mass (COM) is indicated. The system behaves as a pendulum of length $l$ (from pivot to COM). (b) Simulated data for the observed vertical component of force at the pivot during bag filling.
} \label{Fig:schematic}
\end{figure}
The system is observed via the vertical component of the force on the measuring device from which the bag is attached, which is given by:
\begin{equation}
F = m\cos\theta\left(l\omega^2 + g\cos\theta\right).\label{Eq:Force}
\end{equation}
We simulate the system at discrete times $t_n$ to generate some example data. We propagate the dynamical variables between each time according to Eqs.\ (\ref{Eq:theta}) and (\ref{Eq:omega}), and at each timestep evolve the flow of mass into the bag according to:
\begin{equation}
\log\left(\dot{m}_n/1\mathrm{kg\ s^{-1}}\right) = \log\left(\dot{m}_{n-1}/1\mathrm{kg\ s^{-1}}\right) + d_n, \label{Eq:mass_rate}
\end{equation}
where $d_n\sim N(0, \Sigma_m)$ is a process noise term, which models the tendency of the filling-rate to slow down and speed up, whilst remaining strictly positive.
Then each force measurement is simulated as:
\begin{equation}
F_n = F(\theta_n, m_n, l_n) + v_n,
\end{equation}
where the measurement noise $v_n\sim N\left(0, \Sigma_F \right)$.

\subsection{Inference methods}\label{Sec:Inference}
To compare different algorithms' performance in estimating the mass of milk powder in a bag during filling we use three different sequential inference algorithms, each either in conjunction with the physics model described in Sec.\ \ref{Sec:Physics}, or without such a model. Each inference model is fed the same set of measurements $\left\{\mathbf{y}_n\right\}$ (in this case equal to the simulated forces $F_n$), and is given an initial state estimate in the form of a prior distribution. When using the physics model, the algorithms estimate a state vector containing the components $\theta$, $\omega$, $\log\left(m/1\mathrm{kg}\right)$, $\log\left(\dot{m}/1\mathrm{kg\ s^{-1}}\right)$, $\log\left(L/\mathrm{1m}\right)$ and $\log\left(\rho A/\mathrm{1kg\ m^{-1}}\right)$. Otherwise, the algorithms estimate $\log\left(m/1\mathrm{kg}\right)$ and $\log\left(\dot{m}/1\mathrm{kg\ s^{-1}}\right)$ only. Note that log variables of some quantities are estimated to ensure strictly positive estimates of those quantities.

Each algorithm assumes the state vector propagates between timesteps using the forward map $\mathbf{f}:\mathbf{x}_n\mapsto\mathbf{x}_{n+1}$. When the physics model is used, this map integrates the equations of motion [Eqs.\ (\ref{Eq:theta}) and (\ref{Eq:omega})], and evolves the mass as dictated by $\dot{m}$. Otherwise, only the mass is evolved. Each algorithm also maps a state to a measurement estimate by an observation map $\mathbf{g}:\mathbf{x}\mapsto\hat{\mathbf{y}}$, which is given by Eq.\ (\ref{Eq:Force}) when the physics model is used, and $F = mg$ otherwise.

We briefly describe the functioning of each algorithm below, and provide detailed description in Figs.\ \ref{Alg:UKF}-\ref{Alg:PF}.

\subsubsection{Kalman filters}

Firstly, we use two sequential-inference algorithms based on the Kalman Filter, namely the augmented-state Unscented Kalman Filter (UKF), developed in Ref.\ \cite{Julier:2004, Wan:2000}, and the non-augmented UKF, described in Ref.\ \cite{Wan:2001}. These algorithms are extensions of the Kalman Filter designed to perform well with nonlinear forward maps and/or nonlinear observation maps, on the principle that it is easier to approximate the probability distribution than to approximate (linearise) the nonlinear functions. The ability to perform well using nonlinear maps is vital for the current problem, since the equations of motion of the physics model are nonlinear. Even if the physics model is not used, the use of log variables to ensure positive estimates of the mass and mass-flow rate requires the forward map for these variables also to be nonlinear.

Both algorithms approximate the state distribution by specially chosen `sigma points', which capture at least the first two moments of the distribution. The sigma points are run through the forward map and the transformed mean and covariance are used in order to perform a usual Kalman Filter update \cite{Julier:2004}.

\begin{figure}%[!t]
 \removelatexerror
  \begin{algorithm}[H]
  \textbf{Initialise} Set the prior mean $\mu_0$ and covariance $\mathbf{K}_0$ with a well-motivated estimate, along with parameters $\alpha$, $\beta$ and $\gamma$ which determine the distribution of sigma-points \cite{Wan:2000}. Provide the estimated process and measurement covariances: $\Sigma_d$ and $\Sigma_{\nu}$.

\For {$n =1...n_t-1$ }
{
 1) Calculate sigma points and weights $\left\{\mathbf{x}^{(j)}, W^{(j)}\right\}$:\

\Indp 
$\displaystyle \mathbf{x}^{(0)} = \mathbf{\mu}_{n-1}$, \\
$\displaystyle W^{(0)}_m = \frac{1}{N_x +\lambda}$,
$\displaystyle W^{(0)}_c =  \displaystyle W^{(0)}_m + \left( 1 - \alpha^2 + \beta \right)$,

$\displaystyle \mathbf{x}^{(i)} = \mathbf{\mu}_{n-1} + \sqrt{\left(N_x + \lambda\right) \mathbf{K}_{n-1}}$,
$\displaystyle \mathbf{x}^{(i+N_x)} = \mathbf{\mu}_{n-1} - \sqrt{\left(N_x + \lambda\right) \mathbf{K}_{n-1}}$,
$\displaystyle W^{(i)}_m =W^{(i)}_c  = W^{(i+N_x)}_m = W^{(i+N_x)}_c = \frac{1}{2\left(N_x + \lambda \right)}$,

for $i=1...N_x$, where $\lambda = \alpha^2\left(N_x + \kappa \right)$, and $N_x$ is the state dimension.

\Indm 2) Transform sigma points using the forward map:\

\Indp $\displaystyle \hat{\mathbf{x}}^{(i)}_{n} = \mathbf{f}\left( \mathbf{x}^{(i)}_{n}\right)$.

\Indm 3) Calculate the predicted mean and covariance:\

\Indp $\displaystyle \mathbf{\hat{\mu}}_{n} = \sum_{i=0}^{p}W^{(i)}_m\mathbf{\hat{x}}^{(i)}_{n}$,

$\displaystyle \mathbf{\hat{K}}_{n} = \Sigma_d +  \sum_{i=0}^{p} W^{(i)}_c \left( \hat{\mathbf{x}}^{(i)}_{n} -  \mathbf{\hat{\mu}}_{n} \right)\left( \hat{\mathbf{x}}^{(i)}_{n} -  \mathbf{\hat{\mu}}_{n} \right)^T$.

\Indm 4) Recalculate the sigma points using the predicted mean and covariance:\

\Indp 
$\displaystyle \mathbf{\tilde x}^{(0)} = \mathbf{\hat{\mu}}_n$, 

$\displaystyle \mathbf{\tilde x}^{(i)} = \mathbf{\hat{\mu}}_n +  \sqrt{\left(N_x + \lambda\right)  \mathbf{\hat{K}}_{n}}$,

$\displaystyle \mathbf{\tilde x}^{(i+N_x)} = \mathbf{\hat{\mu}}_n - \sqrt{\left(N_x + \lambda\right)  \mathbf{\hat{K}}_{n}}$.

\Indm 5) Apply the  observation model to each new sigma point :\

\Indp $\displaystyle \mathbf{\hat{y}}^{(i)}_{n} = \mathbf{g}\left(\tilde{\mathbf{x}}^{(i)}_{n}\right)$.

\Indm 6) Calculate the predicted observation:\

\Indp $\displaystyle \mathbf{\hat{y}}_n = \sum_{i=0}^{2N_x}W^{(i)}_m \mathbf{\hat{y}}^{(i)}_{n}$.

\Indm 7) Calculate the innovation covariance:\

\Indp $\displaystyle \mathbf{\hat{S}}_n = \Sigma_{\nu} + \sum_{i=0}^{2N_x} W^{(i)}_c \left(\mathbf{\hat{y}}^{(i)}_{n} - \mathbf{\hat{y}}_n \right)\left(\mathbf{\hat{y}}^{(i)}_{n} - \mathbf{\hat{y}}_n \right)^T$.

\Indm 8) Calculate the cross covariance:\

\Indp $\displaystyle \mathbf{K}_{n}^{xy}= \sum_{i=0}^{2N_x} W^{(i)}_c \left( \hat{\mathbf{x}}^{(i)}_{n} - \mathbf{\hat{\mu}}_{n} \right) \left( \hat{\mathbf{x}}^{(i)}_{n} - \mathbf{\hat{\mu}}_{n} \right)^T$.

\Indm 9) Perform a usual Kalman Filter update :\

\Indp $\displaystyle \mathbf{\mu}_{n} = \mathbf{\hat{\mu}}_{n} + \mathbf{W}_n \mathbf{\nu}_n$,

$\displaystyle \mathbf{K}_{n} = \hat{\mathbf{K}}_{n} - \mathbf{W}_n \mathbf{\hat{S}}_n \mathbf{W}_n^T$,

where
$\mathbf{\nu}_n = \mathbf{y}_n - \hat{\mathbf{y}}_n$ and $\mathbf{W}_n = \mathbf{K}_{n}^{xy}\mathbf{\hat{S}}_n^{-1}$.
} 
\end{algorithm}
\caption{Non-augmented-state UKF \cite{Wan:2001}}  \label{Alg:UKF}
\end{figure}

\begin{figure}%[!t]
 \removelatexerror
  \begin{algorithm}[H]
  \textbf{Initialise} Set the prior mean $\mu_0$ and covariance $\mathbf{K}_0$ with a well-motivated estimate, along with parameters $\alpha$, $\beta$ and $\gamma$ which determine the distribution of sigma-points \cite{Wan:2000}. Provide the estimated process and measurement covariances: $\Sigma_d$ and $\Sigma_{\nu}$.

\For {$n =1...n_t-1$ }
{
1) Augment the state mean $\mathbf{\mu}_{n-1}$ and covariance $\mathbf{K}_{n-1}$ with the the process noise  and measurement noise means (zeros) and their respective covariances, $\mathbf{\Sigma}_{d},  \mathbf{\Sigma}_{v}$:

\Indp	
$\mathbf{\mu}_n^a =$
$\left(
\begin{array}{c}
\mathbf{\mu}_{n-1}\\
\mathbf{0}\\
\mathbf{0}\\
\end{array}
\right), $\
$\mathbf{K}_{a,n}= \mathrm{diag}\left(\mathbf{K}_{n-1}, \mathbf{\Sigma}_{d},  \mathbf{\Sigma}_{v}\right)$. \

\Indm 2) Calculate sigma points and weights $\left\{\mathbf{x}^{(j)}, W^{(j)}\right\}$:\

\Indp 
$\displaystyle \mathbf{x}^{(0)} = \mathbf{\mu}_n^a$, \\
$\displaystyle W^{(0)}_m = \frac{1}{N_x +\lambda}$,
$\displaystyle W^{(0)}_c =  \displaystyle W^{(0)}_m + \left( 1 - \alpha^2 + \beta \right)$,

$\displaystyle  \mathbf{x}^{(i)} = \mathbf{\mu}_n^a + \sqrt{\left(N_x + \lambda\right) \mathbf{K}_{a,n}}$, 
$\displaystyle  \mathbf{x}^{(i+N_x)} = \mathbf{\mu}_n^a -  \sqrt{\left(N_x + \lambda\right) \mathbf{K}_{a,n}}$,
$\displaystyle W^{(i)}_m =W^{(i)}_c  = W^{(i+N_x)}_m = W^{(i+N_x)}_c = \frac{1}{2\left(N_x + \lambda \right)}$,
 
for $i=1...N_x$, where $\lambda = \alpha^2\left(N_x + \kappa \right)$, and $N_x$ is the dimension of the augmented state vector.

\Indm 3) Transform sigma points using the forward map:\

\Indp $\displaystyle \hat{\mathbf{x}}^{(i)}_{a,n} = \mathbf{f}_a\left( \mathbf{x}^{(i)}_{a,n}\right)$,
where $\mathbf{f}_a\left(\mathbf{x}_a\right) =$
$\left(
\begin{array}{c}
 \mathbf{f}\left(\mathbf{x}\right) + \mathbf{x}_d\\
\mathbf{x}_d\\
\mathbf{x}_v\\
\end{array}
\right), $\
 and $\mathbf{x}$, $\mathbf{x}_d$ and $\mathbf{x}_v$ are the state, process-noise and measurement-noise parts of $\mathbf{x}_a$.

\Indm 4) Calculate the predicted mean and covariance:\

\Indp $\displaystyle \mathbf{\hat{\mu}}_{a,n} = \sum_{i=0}^{2N_x}W^{(i)}\mathbf{\hat{x}}^{(i)}_{a,n}$,

$\displaystyle \mathbf{\hat{K}}_{a,n} = \sum_{i=0}^{2N_x}\left( \hat{\mathbf{x}}^{(i)}_{a,n} -  \mathbf{\hat{\mu}}_{a,n} \right)\left( \hat{\mathbf{x}}^{(i)}_{a,n} -  \mathbf{\hat{\mu}}_{a,n} \right)^T$.

\Indm 5) Apply the  observation model to each transformed sigma point :\

\Indp $\displaystyle \mathbf{\hat{y}}^{(i)}_{n} = \mathbf{g}_a\left(\hat{\mathbf{x}}^{(i)}_{a,n}\right)$, 
where $\mathbf{g}_a\left(\mathbf{x}_{a}\right) = \mathbf{g}\left(\mathbf{x}\right) + \mathbf{x}_{\nu}$.%, and $\mathbf{x}$ and $\mathbf{x}_{\nu}$ are the state and measurement-noise parts of the augmented state vector $\mathbf{x}_a$.

\Indm 6) Calculate the predicted observation:\

\Indp $\displaystyle \mathbf{\hat{y}}_n = \sum_{i=0}^{2N_x}W^{(i)}_m \mathbf{\hat{y}}^{(i)}_{n}$.

\Indm 7) Calculate the innovation covariance:\

\Indp $\displaystyle \mathbf{\hat{S}}_n = \sum_{i=0}^{2N_x} W^{(i)}_c \left(\mathbf{\hat{y}}^{(i)}_{n} - \mathbf{\hat{y}}_n \right)\left(\mathbf{\hat{y}}^{(i)}_{n} - \mathbf{\hat{y}}_n \right)^T$.

\Indm 8) Calculate the cross covariance (for the non-augmented state):\

\Indp $\displaystyle \mathbf{K}_{n}^{xy}= \sum_{i=0}^{2N_x} W^{(i)} \left( \hat{\mathbf{x}}^{(i)}_{n} - \mathbf{\hat{\mu}}_{n} \right) \left( \hat{\mathbf{x}}^{(i)}_{n} - \mathbf{\hat{\mu}}_{n} \right)^T$.

\Indm 9) Perform a usual Kalman Filter update :\

\Indp $\displaystyle \mathbf{\mu}_{n} = \mathbf{\hat{\mu}}_{n} + \mathbf{W}_n \mathbf{\nu}_n$,

$\displaystyle \mathbf{K}_{n} = \hat{\mathbf{K}}_{n} - \mathbf{W}_n \mathbf{\hat{S}}_n \mathbf{W}_n^T$,

where
$\mathbf{\nu}_n = \mathbf{y}_n - \hat{\mathbf{y}}_n$ and $\mathbf{W}_n = \mathbf{K}_{n}^{xy}\mathbf{\hat{S}}_n^{-1}$.
}
\end{algorithm}
\caption{Augmented-state UKF \cite{Julier:2004, Wan:2000}}  \label{Alg:AUKF}
\end{figure}

\begin{figure}%[!t]
 \removelatexerror
  \begin{algorithm}[H]
  \textbf{Initialise} Sample $n_p$ `particles' from well-motivated prior:

\Indp $\displaystyle \mathbf{x}_{0}^{(i)} \sim P\left(\mathbf{x} \right)$.

Select initial weights $W_{0}^{(i)}=1/n_p$.

Provide the estimated process and measurement covariances: $\Sigma_d$ and $\Sigma_{\nu}$.

\Indm \For {$n =1...n_t-1$ }
{
1) Get next state sample:\

\Indp Propagate `particles' through forward map, adding process noise sampled from relevant distribution (in this case, gaussian)

$\displaystyle  \mathbf{x}_{n}^{(i)} \sim N\left( \mathbf{f}\left(\mathbf{x}_{n-1}^{(i)}\right), \Sigma_d \right)$.

\Indm 2) Get log measurement probability:

\Indp
$\displaystyle \log p\left(\mathbf{y}_n | \mathbf{x}^{(i)}_{n} \right) = -\frac{1}{2} \mathbf{\nu}_{i}^T \Sigma_{\nu}^{-1}\mathbf{\nu}_{i} $, where $\mathbf{\nu}_i = \mathbf{y}_n - \mathbf{g}\left(\mathbf{x}^{(i)}_{n} \right)$

and set weights
$\displaystyle  \hat{W}_{n}^{(i)} = \exp\left(\log W_{n-1}^{(i)} + \log p\left(\mathbf{y}_n | \mathbf{x}_{n}^{(i)} \right) \right)$.

Normalise weights:
$\displaystyle W_{n}^{(i)} = \frac{\hat{W}_{n}^{(i)}}{\sum_j \hat{W}_{n}^{(j)}}$.

\Indm 3) Get effective number of particles:

\Indp $\displaystyle n_{\mbox{\scriptsize eff}} = \frac{1}{ \sum_j \left(\hat{W}_{n}^{(i)}\right)^2   }$.

\Indm

\If {$n_{\mbox{\scriptsize eff}}<n_{\mbox{\scriptsize thr}}$}
{
Draw $n_p$ particles $\left\{x_{n}^{(j)}\right\}$ from the current particle set $\left\{x_{n}^{(i)} \right\}$ with probabilities proportional to $\left\{W_{n}^{(i)}\right\}$.

Reset weights 
$\displaystyle W_{n}^{(j)} = 1/n_p$.
}
}
\end{algorithm}
\caption{Particle filter}  \label{Alg:PF}
\end{figure}

%\begin{center}
%\begin{minipage}{\columnwidth}
%  \centering  
%  \myparticlealgorithm
%\end{minipage}%
%\captionof{figure}{Particle filter}  \label{Alg:PF}
%\end{center}
The augmented-state UKF differs from the non-augmented UKF in 3 ways: in the augmented-state UKF the sigma-point states are augmented with parameters representing the process noise and measurement noise; the non-augmented state UKF samples the sigma-points twice, while the augmented-state UKF only once; also, in the non-augmented state UKF the process and measurement uncertainties are included in the calculation through the `augmented maps' $\mathbf{f}_a$ and $\mathbf{g}_a$ (see Fig.\ \ref{Alg:AUKF}) rather than added to the predicted covariance and innovation covariance as in the non-augmented UKF (see Fig.\ \ref{Alg:UKF}).

 The performance of these two algorithms is compared in Ref.\ \cite{Wu:2005}, which found that for the problems considered there  the augmented-state UKF generally performs better.

%%%%%%%%%%%%%

%\begin{algorithm}
%\caption{EuclidÕs algorithm}\label{alg:euclid}
%\begin{algorithmic}[1]
%\Procedure{Euclid}{$a,b$}\Comment{The g.c.d. of a and b}
%\State $r\gets a\bmod b$
%\While{$r\not=0$}\Comment{We have the answer if r is 0}
%\State $a\gets b$
%\State $b\gets r$
%\State $r\gets a\bmod b$
%\EndWhile\label{euclidendwhile}
%\State \textbf{return} $b$\Comment{The gcd is b}
%\EndProcedure
%\end{algorithmic}
%\end{algorithm}
\subsubsection{Particle filters}
As well as the two varieties of UKF, we also test the performance of a particle filter. We use a sequential importance resampling algorithm, first introduced in Ref. \cite{Gordon:1993}. This algorithm has the advantage that as well as permitting a nonlinear forward map and observation function, it permits the prior distribution for the state vector to have any form, as well as the estimated noise distributions (although we do not need to exploit this flexibility for the current problem).
The algorithm approximates a distribution of states by a sample of weighted states (`particles'). It evolves each `particle' through the forward map before performing a Bayesian update of the sample. Resampling prevents weights becoming concentrated in a small number of particles.

\section{Results}
We ran the algorithms on measurements simulated using the following parameters/initial conditions (partly motivated by Ref.\ \cite{Tuohy:1989}): 
$\theta_0=0.2$ rad, $\omega_0=0.2$ rad/s, $L=3.5$ m, $m_0=1.7$ kg, $\dot{m}_0 = 5.5$ kg/s, $\rho A = 161.25$  kg/m. We used $\Sigma_m = 0.1$ to generate process noise in the mass flow [via Eq.\ (\ref{Eq:mass_rate})] at samples every 0.025s during 10s of filling time.
Both UKF algorithms used a prior distribution for the state space with mean values equivalent to: $\theta_0=0.21$ rad, $\omega_0=0.15$ rad/s, $L=2.5$ m, $m_0=2.2$ kg, $\dot{m}_0 = 5.36$ kg/s, $\rho A = 177.38$  kg/m and variances: $\sigma^2_\theta$ = 0.2 rad$^2$, $\sigma^2_\omega$ = 0.2 rad$^2$, 
$\sigma^2_{\log{L/1\mathrm{m}}}$ = 0.2, $\sigma^2_{\log{m/1\mathrm{kg}}} = $ 0.02, $\sigma^2_{\log{\dot{m}/1\mathrm{kg\ s^{-1}}}} = $ 0.02, and $\sigma^2_{\log{\rho A/1\mathrm{kg\ m^{-1}}}} = 0.2$. This prior overlapped the `true' state, but was not centred thereon. The particle filter sampled 1000 particles from a gaussian prior of the same mean and covariance as used by the UKF algorithms. For the estimated process noise covariance, $\Sigma_d$, we used $\Sigma_m$ for the $\log\left(\dot{m}/1\mathrm{kg\ s^{-1}}\right)$ variance component, and small but non-zero values for the other diagonal components. These non-zero values for the variables without expected process noise are necessary for the correct functioning of all the algorithms, and are usually justified in real applications by invoking the need to account for the difference between the (necessarily simplistic) model used and the laws governing the true system dynamics. In the simulated measurements, the noise covariance $\Sigma_F = 1.5\ \mathrm{N}^2$, and the estimated noise covariance supplied to the inference algorithms $\Sigma_{\nu}=2.5\ \mathrm{N}^2$
\begin{figure}
%\centering
\includegraphics[width=\columnwidth]{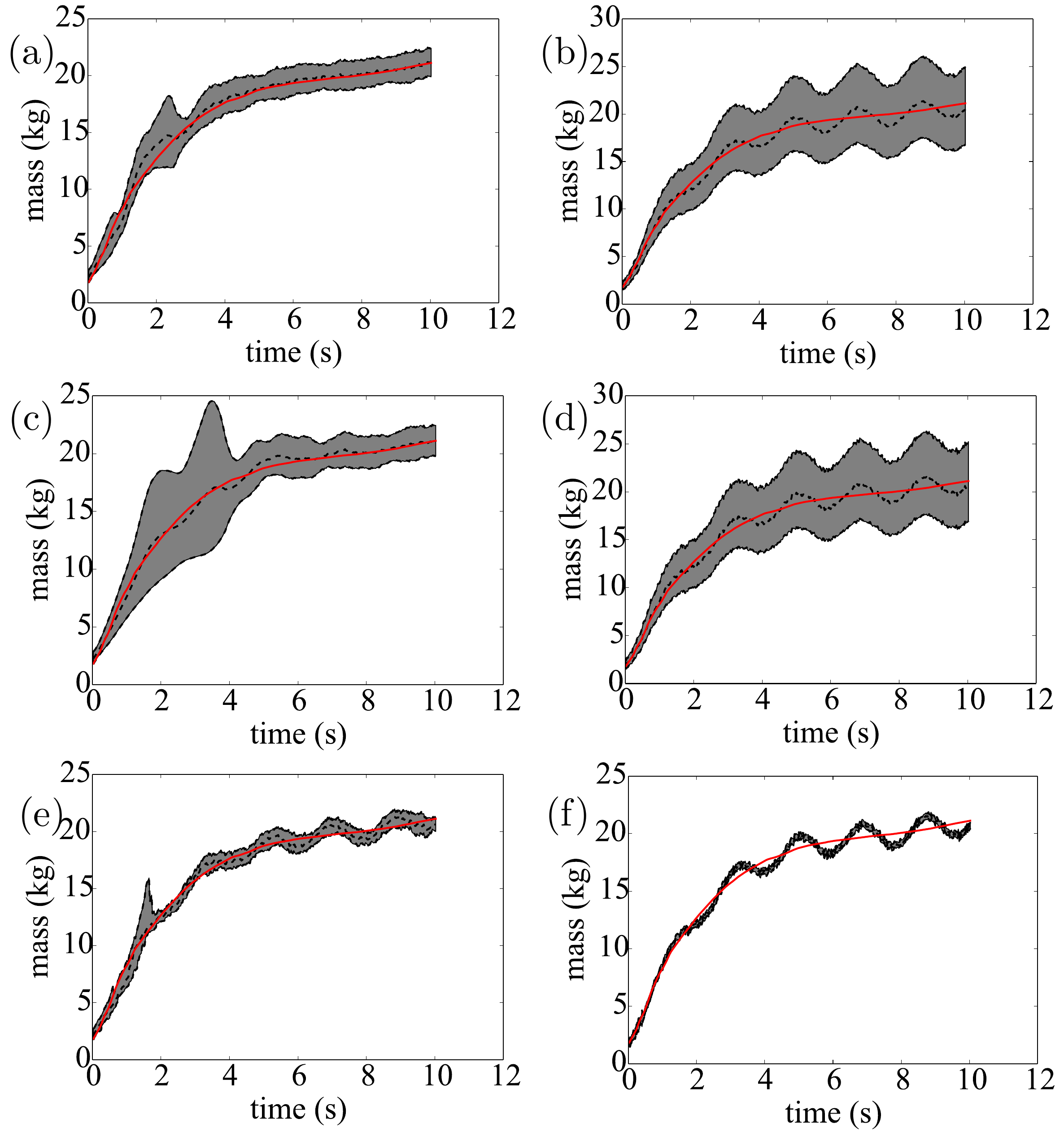}
\caption{Sequential mass estimates (shaded regions) and true mass values [light (red) lines] as a function of time. Subplots (a) and (b) show results for the non-augmented-state UKF, (c) and (d) for the augmented-state UKF, and (e) and (f) for the particle filter. Results shown in (a), (c) and (e) are computed using the physics model and those in (b), (d) and (f) without. For the UKF results, grey regions represent the central $90\%$ prediction regions for the mass, dotted lines represent the region boundaries and mean mass prediction; for the particle filter, grey regions represent masses between the 5th and 95th percentile of the particle sample (indicated by dotted lines), the 50th mass percentile is also given by a dotted line.
} \label{Fig:masses}
\end{figure}

\begin{figure}
%\centering
\includegraphics[width=\columnwidth]{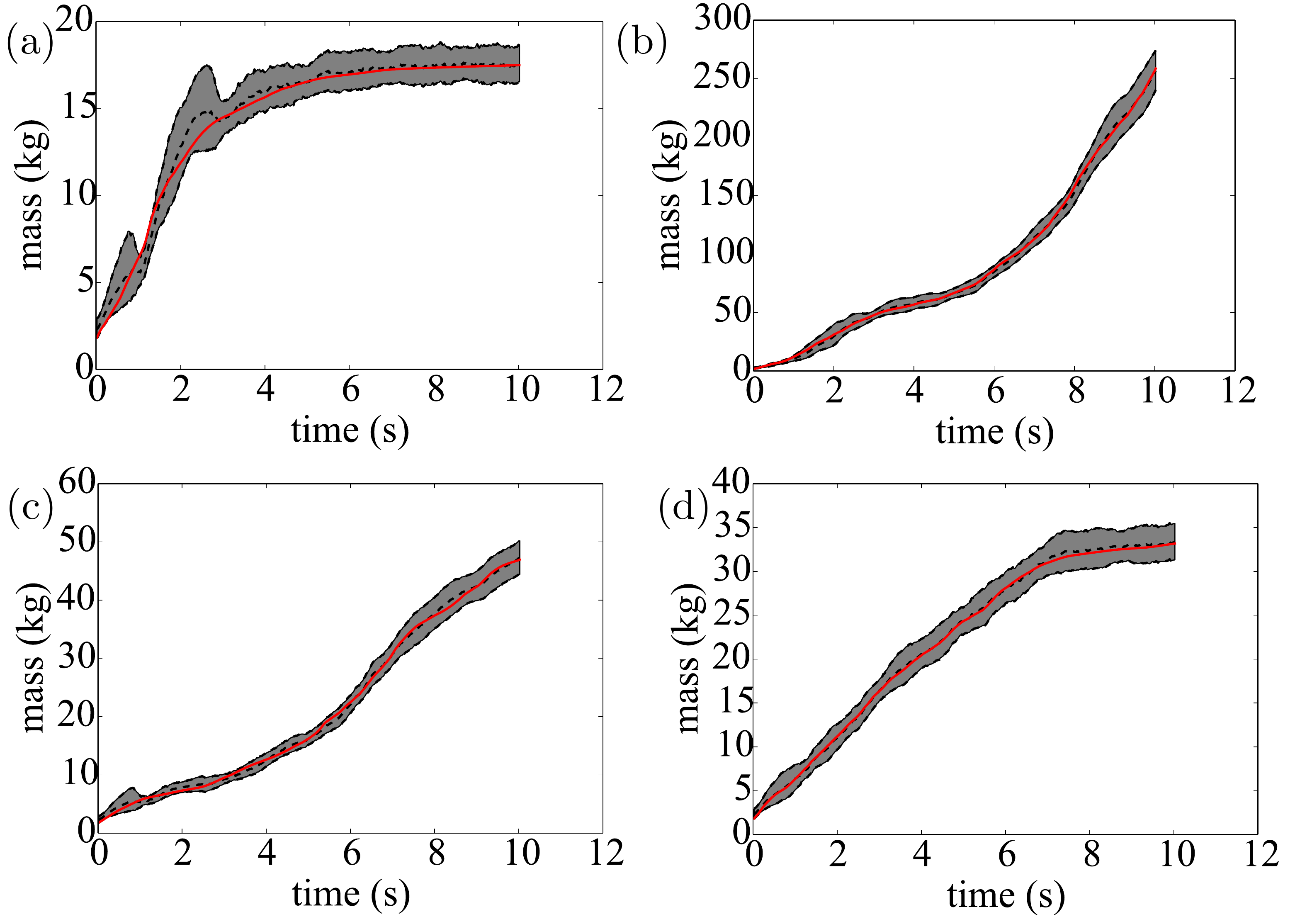}
\caption{Sequential mass estimates (shaded regions) from for non-augmented-state UKF and true mass values [light (red) lines] as a function of time. Grey regions represent the central $90\%$ prediction regions for the mass, dotted lines represent the region boundaries and mean mass prediction. Plots (a)-(d) show results obtained for simulated data with different realisations of the process and measurement noise.
} \label{Fig:realisations}
\end{figure}

We ran all algorithms on the same set of simulated measurements illustrated in Fig.\ \ref{Fig:schematic}(b).  The sequential estimates of mass are expressed as $90\%$ prediction regions, and are shown in Fig.\ \ref{Fig:masses}. The most successful methods were the UKFs and particle-filter using the physics model [Figs.\ \ref{Fig:masses}(a), (c) and (e)]. These  methods produced suitably narrow regions which contained the `true' mass value. The UKFs which did not use a physics model provided wide $90\%$ prediction intervals which contained the true mass value [Figs.\ \ref{Fig:masses}(b) and (d)]; the particle filter which did not use a physics model produced a narrow $90\%$ prediction interval which often did not contain the true mass value [Fig.\ \ref{Fig:masses}(f)]. The particle filter without a model clearly attributed the oscillations in the measured vertical force component to oscillations in the mass flow, even though the magnitude of these oscillations was larger than would be expected from the stated process noise. We tested the non-augmented UKF for different realisations of the process and measurement noise, and it was found to produce consistently good estimates [see Fig.\ \ref{Fig:realisations}].

\begin{figure}
%\centering
\includegraphics[width=\columnwidth]{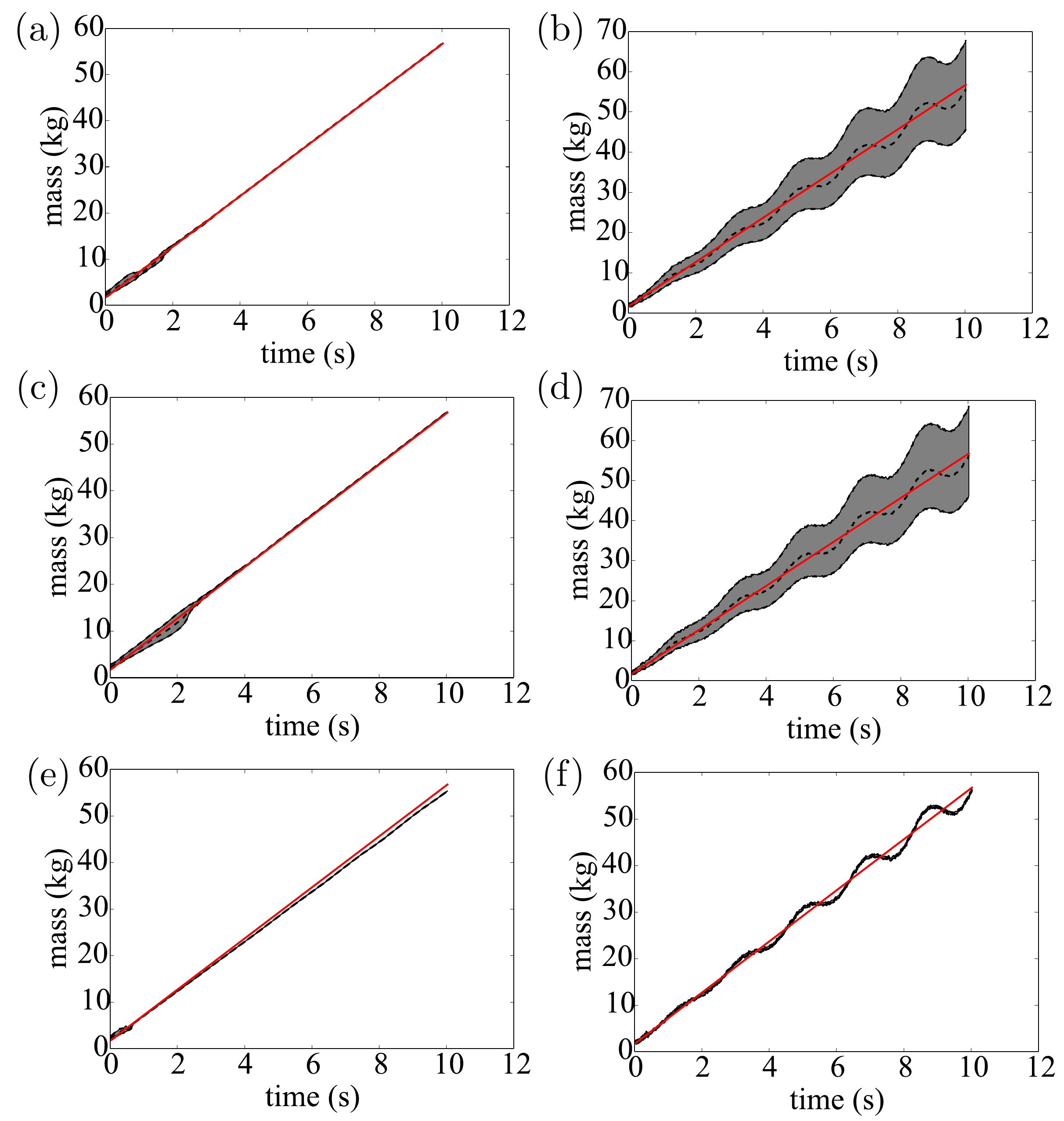}
\caption{As Fig.\ \ref{Fig:masses}, but all with all inferences performed on simulated data with negligible process noise in $\dot{m}$, using negligible estimated process noise.
} \label{Fig:masses2}
\end{figure}

As time $t\rightarrow \infty$, the widths of the prediction intervals are limited by the effect of the finite process and measurement noise. By using idealised equipment, where the process and measurement noise is very small, it should be possible to obtain very accurate estimates of the mass. We demonstrate this by simulating a force dataset using negligible process noise in $\dot{m}$. We ran all inference algorithms on this dataset using estimates of the process covariance $\sigma_d$ with very small diagonal elements of order $10^{-8}$ in each appropriate unit (and zero-valued off-diagonal elements). All inference methods using the physics model produced estimates with very narrow prediction region [Fig.\ \ref{Fig:masses2}(a), (c) and (e)]; however, the particle filter's 90\% prediction region did not contain the true mass value. For these simulations with no process noise [Figs.\ \ref{Fig:masses2} (b), (d) and (f)], the methods without a physics model produced poor estimates, either with wide prediction regions (in the case of the UKFs), or with narrow regions not always containing the true mass value (in the case of the particle filter).

\section{Conclusions}
We have shown the benefit of including a physics model in sequential inference algorithms used in control-systems for automated weighing - in particular for weighing a suspended bag of powdered milk during the filling process. When used in conjunction with a UKF or particle filter algorithm, these methods are expected to produce narrow $90\%$ prediction regions containing the true mass value, which would not be possible without the consideration of the dynamics generated by the physics model. Reference \cite{Wu:2005} found that the augmented-state UKF performed better than the non-augmented UKF in the problems considered there; however, for the problem considered in this paper, both UKF algorithms produce very similar output in similar running times. UKF algorithms are expected to be more useful than particle filters, since their running time is comparable to the filling times used in this paper. The particle filter's running time is several times longer than these filling times, so in practice, particle filters would require slower filling rates and longer times between force measurements in order to inform a synchronous control system. Improved predictive performance of the particle filter in the cases where the prediction intervals fail to contain the true mass value would be expected if a larger number of particles were sampled. However, this would slow the running time even further.

Consideration of systems with process noise in the density $\rho$ and cross-sectional area $A$ would provide straightforward and important extensions of this work. The density of milk would be expected to fluctuate as it settles during packing, and the cross-sectional area $A$ would also be expected to vary with height within the bag.

\section*{Acknowledgment}
This work was funded by grant UOOX1208 from the Ministry of Business, Innovation \& Employment.



\bibliographystyle{IEEEtran}
\bibliography{IEEEabrv,./variable_pendulum}
%\begin{thebibliography}{1}
%\end{thebibliography}

\end{document}